# Solution to the Equity Premium Puzzle

Atilla Aras

November 10, 2020

**Funding**: This research did not receive any specific grant from funding agencies in the public, commercial, or not-for-profit sectors.

# Solution to the Equity Premium Puzzle

## Abstract

This study provides a solution of the equity premium puzzle. Questioning the validity of the Arrow-Pratt measure of relative risk aversion for detecting the risk behavior of investors under all conditions, a new tool, that is, the sufficiency factor of the model was developed to analyze the risk behavior of investors. The calculations of this newly tested model show that the value of the coefficient of relative risk aversion is 1.033526 by assuming the value of the subjective time discount factor as 0.99. Since these values are compatible with the existing empirical studies, they confirm the validity of the newly derived model that provides a solution to the equity premium puzzle.

## 1. Introduction

The equity premium puzzle is a quantitative puzzle that implies the inability of intertemporal economic models to explain the large historical equity premium under reasonable parameter values in the US financial markets over the past century. The equity premium puzzle, coined by Mehar and Prescott (1985), arises because high historical equity premium leads to an unreasonably high level of risk aversion among investors according to standard intertemporal economic models. The puzzle is important for both academic world and practitioners because intertemporal economic models do not replicate such a high historical equity premium under reasonable parameters. In this study, a solution for this unsolved puzzle has been proposed.

The objectives of this study are to give a solution to the equity premium puzzle and to question the validity of the Arrow-Pratt measure of relative risk aversion for detecting the risk behavior of investors under all conditions.

The motivation of the study depends on the implicit assumption of the fair gamble that risk behavior of individuals can be expressed by means of uncertain utility curves.

Individuals decide about an uncertain value by comparing its utility with that of a certain value. The behavior of individuals toward risk, as a risk-averse, risk-loving, or risk-neutral decision maker, can be observed at the time the individuals compare the utilities of certain and uncertain wealth values. Since individuals allocate utility or negative utility for the uncertain wealth value at the time of the comparison, definitions of risk-averse, risk-loving or risk-neutral investor have been reformulated in this study.

The assumptions of the coefficient of relative risk aversion under fair gamble, that is, the assumptions of $\rho = \frac{2\pi w_s}{\sigma_{\hat{z}}^2}$ have been questioned. A new formula for the coefficient of relative risk aversion has been derived under assumptions that are different from those of fair gamble. Since the derived formula for the coefficient of relative risk aversion has three components that are different from $\rho = \frac{2\pi w_s}{\sigma_{\hat{z}}^2}$, it has been concluded that the coefficient of relative risk aversion may not imply risk behavior of individuals under all conditions and its magnitude may not be the appropriate coefficient to discriminate between risk-averse, risk-loving, and risk-neutral investors again under all conditions. Hence, a new tool, that is, the sufficiency factor of the model, has been developed to classify the investors as risk-averse, risk-loving, and risk-neutral.

The problem of a typical investor has been reformulated in the new model. Following Mehra (2008), Cochrane (2009), and Danthine and Donaldson (2014), a new system of equations has been developed.

The values of the coefficient of relative risk aversion and the sufficiency factor of the model for the investors who invested in equity and risk-free asset in US economy for the 1889-1978 period have been calculated from the new system of equations.

The coefficient of relative risk aversion and the subjective time discount factor are 1.033526 and 0.99, respectively. Since these values are compatible with the empirical studies, they confirm that the new model gives a solution to the equity premium puzzle.

## 2. Materials and methods

The standard intertemporal economic models do not explain that the large historical equity premium leads an implausibly high level of risk aversion in US financial markets in the past century. This is the equity premium puzzle that is coined by Mehra and Prescott (1985).

Mehra and Prescott coined the equity premium puzzle by arguing that consumption capital asset pricing model (CCAPM) is unable to explain the high historical US equity premium over the past century under reasonable parameter values. CCAPM is an economic model that provides a theoretical connection between aggregate consumption and financial markets (Danthine & Donaldson, 2014). CCAPM indicates that the representative agent's utility function, his or her subjective time discount factor, and the process on consumption that equals output or dividends in the exchange economy equilibrium are the only factors that determine the security returns (Danthine & Donaldson, 2014). Up till now, no agreed solution for the puzzle has been provided so that CCAPM replicates this high equity premium under reasonable parameter values.

The studies that propose solution to the equity premium puzzle can be grouped into seven classes (Mehra 2003; Mehra 2008).

The first class consists of existing literature that presents preference modifications for the solution (Abel, 1990; Benartzi & Thaler, 1995; Campbell & Cochrane, 1999; Costantinides, 1990; Epstein & Zin, 1991).

The second class consists of studies that present survival bias for the solution (Brown et al., 1995).

The third class is based on studies that present market imperfections for the solution (Aiyagari & Gertler, 1991; Alvarez & Jermann, 2000; Bansal & Coleman, 1996; Costantinides et al., 2002; Heaton & Lucas, 1996; McGrattan & Prescott, 2001).

The fourth class consists of studies that present modified probability distributions to admit rare but disastrous events for the solution (Rietz, 1988).

The fifth class is based on existing literature that presents incomplete markets for the solution (Costantinides & Duffie, 1996; Heaton & Lucas, 1997; Mankiw, 1986).

The sixth class comprises studies that present limited participation of consumers in the stock market for the solution (Attanasio et al., 2002; Brav et al., 2002).

The seventh class consists of studies that present temporal aggregation problem for the solution (Gabaix & Laibson, 2001; Heaton, 1995; Lynch, 1996).

Individuals choose between certain and uncertain wealth values by comparing their utilities stemmed from these values. None of the proposed solutions take into account that individuals may increase or decrease their utility of uncertain wealth value according to their risk preferences at the time of the above-mentioned comparison. This observation of the author of the article will be incorporated into the new model in subsequent sections.

## 3. Theory

### 3.1 Criticism of the Arrow-Pratt Coefficient of Relative Risk Aversion

Many micro literature proposes that coefficient of relative risk aversion must be approximately equal to 1. Nevertheless, some economists believe that it can take values as high as 2 or 4. Despite all this, there seems to be consensus among economists that this coefficient has to be lower than 10.

In finance theory, when investors make a decision about the uncertain wealth value, the standard method is to design a lottery and compare the expected value of its utility to the utility of certain wealth value. The expected value of the lottery is estimated to determine the payoff of the lottery that is a future variable. However, the payoff of the lottery determined by this method may be incorrect due to future uncertainty or insufficiency of the method used. Since it is impossible to predict the future values correctly most of the time by any method, the investor will not trust that method completely. Therefore, it is a necessity to reflect future uncertainty and insufficiency of the method used on the uncertain utility gained from the future values at the time individuals compare the utilities of certain and uncertain values. Thus, the investor may increase or decrease his uncertain utility by taking into account this future uncertainty and insufficiency of the method according to his/her risk preference. Hence, a related parameter must be reflected in the uncertain utility gained from uncertain values at the time of comparison. This reflection in the uncertain utility takes the form of risk behavior of the investor.

In the derivation of $\rho = \frac{2\pi w_s}{\sigma_{\hat{z}}^2}$, the following points are unclear: As shown in Figure 1, it is assumed that the utility of the certain wealth value and the utility of the uncertain wealth value at ($w$-$\Delta$) and ($w$+$\Delta$) wealth values intersect. [Insert figure 1 here] However, it is assumed that the utility of the certain wealth value and the utility of the uncertain wealth value do not intersect at w. Moreover, it is also assumed that the utility curve of the uncertain wealth value is a straight line between $u(w$-$\Delta)$ and $u(w$+$\Delta)$ required in the derivation of $\rho = \frac{2\pi w_s}{\sigma_{\hat{z}}^2}$. This practice cannot be a general true for examining the risk behavior of people because the utility curves of certain and uncertain wealth values are assumed to intersect at some wealth values (i.e., they are equal at these values) and the utility curve of uncertain wealth value is assumed to be a straight line. This practice cannot be assumed under all conditions. The study further analyzes the situation when the investor has the utility curve of a lottery such as $u_2$ in Figure 2 or has the utility curve of uncertain wealth value in Figure 3. [Insert figure 2 here] [Insert figure 3 here]. The assumption that the utility curve of the uncertain wealth value may be the one in Figure 3 changes the formula of $\rho = -\frac{u''(w)}{u'(w)}w$ in terms of risk premium. Hence, a new fact has been developed. As the mathematical form of the utility curve of the lottery changes, the implications of $\rho = -\frac{u''(w)}{u'(w)}w$ change. This will be shown in the next section.

When the risk behavior of individuals is investigated by $\rho = \frac{2\pi w_s}{\sigma_{\tilde{z}}^2}$, the comparison of the utilities of the certain and uncertain wealth values is carried out at the same wealth value, while the subjective time discount factor for the utility of uncertain wealth value is not used in the comparison. However, there is no need to compare the utilities of the certain and uncertain wealth values at the same wealth value. This is because for instance the risk behavior of risk-averse investors is such that their attitude toward risk may reduce the utility of the uncertain wealth value below the utility of the certain wealth value at the time of comparison. Therefore, the behavior of risk-averse individuals is such that the negative utility allocated for the uncertain wealth level at the time of comparison, reduces the utility of the uncertain wealth level below the utility of the certain wealth level. Hence, the comparison should be done between the utilities of certain and uncertain values at wealth values irrespective of numerical values, but the utility of uncertain value must be discounted by the subjective time discount factor.

Hence, the definitions for risk-averse, risk-loving, and risk-neutral investors are reformulated in this study as follows:

*A risk-averse investor* allocates negative utility or zero utility for uncertain wealth value due to insufficient model and future uncertainties at time t so that

$$u(w_t) > \beta \eta_t u(w_T) \tag{1}$$

holds true. Here, u is a continuously differentiable and increasing concave utility curve; t is time the investor compares the utilities of certain and uncertain wealth values; $w_t$ is the certain wealth value at time t; $w_T$ is a guess at time t of the uncertain future wealth value ($w_{t+1}$) by using the information of time t; $\beta$ is the subjective time discount factor and $\eta_t$ is the *sufficiency factor of the model* that is coined by the author of the article. $\eta_t$ is a coefficient that is determined at time t for the utility curve of the uncertain value, that is, $u(w_T)$. It is calculated as follows:

$\eta_t u(w_T) = u(w_T) +$ negative utility allocated by the investor at time t due to insufficient model and future uncertainties                    (2)

$\eta_t u(w_T) = u(w_T) +$ utility allocated by the investor at time t due to insufficient model and future uncertainties                    (3)

$\eta_t u(w_T) = u(w_T) +$ zero utility allocated by the investor at time t due to insufficient model and future uncertainties                    (4)

$\eta_t$ is determined at time t because the utility or the negative utility due to insufficient model and future uncertainties are allocated by the investor for the uncertain wealth value at time t. Time t denotes the time the investor compares the certain and uncertain utility values. Hence, behavior toward risk as risk-averse, risk-loving or risk-neutral is an activity at time t, that is, risk behavior is observed at time t.

Moreover, the utility of the uncertain wealth value is of the form $\eta_t u(w_T)$ because the utility of the certain value is of the form $u(w_t)$.

If expectation operator is used for the prediction of the uncertain utility gained from future wealth value ($w_{t+1}$) with the information set available at time t, we will have the following for the risk-averse investor:

$$u(w_t) > \beta \eta_t E_t[u(w_{t+1})]$$                    (5)

Moreover, *not enough risk-averse investor* allocates negative utility for uncertain wealth value due to the insufficient model and future uncertainties at time t so that

$$u(w_t) \geq \beta \eta_t u(w_T) \ or \ u(w_t) \geq \beta \eta_t E_t[u(w_{t+1})]$$                    (6)

does not hold true.

A *risk-loving investor*, on the other hand, allocates utility or zero utility for uncertain wealth value due to the insufficient model and future uncertainties at time t so that

$$u(w_t) < \beta\eta_t u(w_T) \; or \; u(w_t) < \beta\eta_t E_t[u(w_{t+1})] \qquad (7)$$

holds true. This is because there is a chance that the investor may profit from the insufficient model and future uncertainties in the future.

*Not enough risk-loving investor* allocates utility for uncertain wealth value due to the insufficient model and future uncertainties at time t so that

$$u(w_t) \leq \beta\eta_t u(w_T) \; or \; u(w_t) \leq \beta\eta_t E_t[u(w_{t+1})] \qquad (8)$$

does not hold true.

Finally, a *risk-neutral investor* allocates utility, negative utility or zero utility for uncertain wealth value due to the insufficient model and future uncertainties so that

$$u(w_t) = \beta\eta_t u(w_T) \; or \; u(w_t) = \beta\eta_t E_t[u(w_{t+1})] \qquad (9)$$

holds true.

An example will be given to make definitions more concrete. The investor will decide to buy or sell his/her asset by taking into account the inequalities below. Assume that the risk-averse investor has a stock that has a price of \$2. The investor expects the price of the stock to increase to \$2.2 in the next period. Because he/she is risk-averse, he/she allocates negative utility to the utility of \$2.2 so that

$$u(\$2) > \beta[u(\$2.2) + \text{negative utility allocation by the investor}]$$

$$= u(\$2) > \beta\eta u(\$2.2) \text{ holds true.}$$

Consider now figures 3 and 4 that demonstrate the increasing and continuously differentiable concave utility curves $u(w)$ for the investor who allocates negative utility for the uncertain wealth value, where $w_s$ denotes the certain wealth value. [Insert figure 4 here] The investor expects $w_s$ to increase to $w_i$, where $w_i$ is the uncertain wealth value. $w_f$ denotes the uncertain wealth value that is larger than $w_i$. CE denotes the certainty equivalent value. $\pi$ is the risk premium. $\delta_t$ is the constant utility value associated with the insufficiency of the model and future uncertainty for all wealth levels. $\eta_t$ is the sufficiency factor of the model.

CE and $\pi$ are defined in the same way as they are defined in $\rho = \frac{2\pi w_s}{\sigma_{\hat{z}}^2}$, that is, CE is the certain wealth value that provides equal utility to the random payoff of the lottery and $\pi$ is the wealth value that a person has to pay in order to not participate in the lottery.

The investor expects the certain value $(w_s)$ to increase to the uncertain value $(w_i)$. In the future, $w_i$ may be smaller than, larger than, or equal to $w_i$. Since $w_i$ is a prediction and the investor does not trust it completely, the utility gained from $w_i$ will be equal to $[u(w_i) - \delta_t]$, where $-\delta_t$ is the constant negative utility value allocated due to the future uncertainty and the insufficiency of the model. It has been noted that $[u(w_i) - \delta_t]$ represents the undiscounted utility of the payoff of the lottery.

Since the investors allocate different values of negative utility to different wealth levels due to the future uncertainty and insufficiency of the models, the undiscounted utility of $w_i$ of the investor will be $\eta_t u(w_i)$, where $\eta$ will take values as follows:

$\eta u(w)$ will be below $u(w)$ if

$$\eta < 1 \text{ and } \rho < 1 \tag{10}$$

and

$$\eta > 1 \text{ and } \rho > 1 \tag{11}$$

hold true.

$\eta u(w)$ will be above $u(w)$ if

$$\eta > 1 \text{ and } \rho < 1 \tag{12}$$

and

$$\eta < 1 \text{ and } \rho > 1 \tag{13}$$

hold true.

It is observed that $\eta_t u(w_i)$ represents the undiscounted utility of the payoff the lottery. The above inequalities have been demonstrated in Figures 5 and 6. [Insert figure 5 here] [Insert figure 6 here]

Following Equations 10 and 11, the investor can allocate different values of negative utility to different wealth levels due to future uncertainty and insufficiency of the model. Since utility is hard to quantify in practice, the utility curve of $[u(w_i) - \delta_t]$ will be less useful mathematically and practically. Moreover, the investor allocates the same value of negative utility to different wealth levels using $[u(w_i) - \delta_t]$. Hence, this allocation is also impractical from the viewpoint of economics.

When the investor expects the certain wealth value $(w_s)$ to fall to the uncertain wealth value $(w_d)$, we observe Figure 7, where $w_p$ denotes the future wealth value that is smaller than $w_d$. [Insert figure 7 here] CE shows the certainty equivalent value and $\pi$ is the risk premium as before.

## 3.2 Magnitude of the Arrow-Pratt Measure of Relative Risk Aversion

Now, consider Figure 3. For the investor, $\pi$ is equal to $[w_s$- CE] according to their standard definitions. As observed, $\pi$ is not equal to $[w_i$- CE] and is therefore different from the one that exists in the $\rho = \frac{2\pi w_s}{\sigma_{\tilde{z}}{}^2}$. Here, $w_s$ and $w_i$ denote certain and uncertain wealth values, respectively. Moreover, $w_i$ is mathematically equal to the sum of $w_s$ and $E\tilde{z}$. In other words,

$$w_i = w_s + E\tilde{z} \qquad (14)$$

Here, $\tilde{z}$ is a lottery as follows:

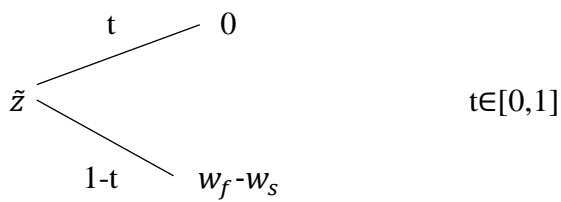

t∈[0,1]

or,

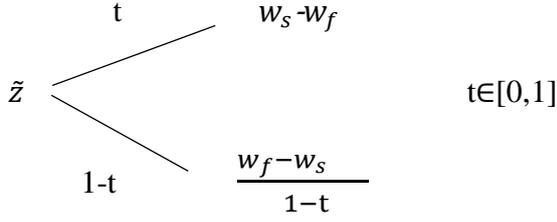

$E\tilde{z}$ is

$$0.t + (1-t)w_f\text{-}w_s = (1-t)w_f\text{-}w_s \tag{15}$$

or,

$$t(w_s\text{-}w_f) + 1\text{-}t\left(\frac{w_f\text{-}w_s}{1-t}\right) = (1-t)w_f\text{-}w_s \tag{16}$$

The investor may not trust $\tilde{z}$ fully because the future realization of $E\tilde{z}$ may be negative due to future uncertainty and insufficiency of the model.

From the properties of the utility curves of the investor from Figure 3, we get:

$$u(w_s - \pi) = \beta u(w_i) - \delta_t = \beta u(w_s + E\tilde{z}) - \delta_t \tag{17}$$

or,

$$u(w_s - \pi) = \beta\eta_t u(w_i) = \beta\eta_t u(w_s + E\tilde{z}) . \tag{18}$$

Expanding $u$ with around $w_s$ on both sides of Equations 17 and 18, the following can be obtained:

$$u(w_s - \pi) = u(w_s) - \pi u'(w_s) + O(\pi^2), \tag{19}$$

$$\beta u(w_s + E\tilde{z}) - \delta_t = \beta u(w_s) + \beta(E\tilde{z})u'(w_s) + \frac{1}{2}\beta(E\tilde{z})^2 u''(w_s) - \delta_t + O[(E\tilde{z})^3], \tag{20}$$

$$\beta\eta_t u(w_s + E\tilde{z}) = \beta\eta_t u(w_s) + \beta\eta_t(E\tilde{z})u'(w_s) + \frac{1}{2}\beta\eta_t(E\tilde{z})^2 u''(w_s) + O[(E\tilde{z})^3] \tag{21}$$

Here, $u'(w_s)$ and $u''(w_s)$ denote the first and second derivative of u with respect to $w_s$, respectively.

Moreover, O denotes the terms of order at the most. Here, the behavior of $\alpha = -\frac{u''(w)}{u'(w)}$ as $(E\tilde{z})^3 \to 0$ will be examined.

If Equations 19 and 20 are substituted in Equation 17

and

Equations 19 and 21 are substituted in Equation 18, the following can be derived:

$$\alpha = \frac{2\pi}{\beta(E\tilde{z})^2} + \frac{2}{E\tilde{z}} - \frac{2\delta_t}{\beta u'(w_s)(E\tilde{z})^2} + \frac{2u(w_s)(\beta-1)}{\beta u'(w_s)(E\tilde{z})^2}, \qquad (22)$$

$$\alpha = \frac{2\pi}{\beta \eta_t (E\tilde{z})^2} + \frac{2}{E\tilde{z}} + \frac{(\beta \eta_t - 1)2u(w_s)}{\beta \eta_t u'(w_s)(E\tilde{z})^2}. \qquad (23)$$

Moreover, $\rho = -\frac{u''(w)}{u'(w)}w$, the sufficiency factor of the model and the risk premium will be as follows:

$$\rho = \frac{2\pi w_s}{\beta(E\tilde{z})^2} + \frac{2w_s}{E\tilde{z}} - \frac{2\delta_t w_s}{\beta u'(w_s)(E\tilde{z})^2} + \frac{2u(w_s)(\beta-1)w_s}{\beta u'(w_s)(E\tilde{z})^2}, \qquad (24)$$

$$\rho = \frac{2\pi w_s}{\beta \eta_t (E\tilde{z})^2} + \frac{2w_s}{E\tilde{z}} + \frac{(\beta \eta_t - 1)2u(w_s)w_s}{\beta \eta_t u'(w_s)(E\tilde{z})^2}, \qquad (25)$$

$$\eta_t = \frac{2\pi w_s u'(w_s) - 2w_s\, u(w_s)}{\rho \beta u'(w_s)(E\tilde{z})^2 - 2w_s E\tilde{z}\beta u'(w_s) - 2w_s \beta\, u(w_s)}, \qquad (26)$$

$$\pi \cong \frac{\rho[\eta_t \beta u(w_{ns}) - u(w_s)]}{u''(w_s)} \qquad (27)$$

(See Derivation of Equation 27 in Appendix for proof)

Here, $w_{ns}$ and $w_s$ denote the uncertain and certain wealth values, respectively.

As observed, $\rho = -\frac{u''(w)}{u'(w)}w$ includes three components in Equation 25 as compared to one component in $\rho = \frac{2\pi w}{\sigma_{\tilde{z}}^2}$. Therefore, the magnitude of $\rho = -\frac{u''(w)}{u'(w)}w$ may not be interpreted in terms of the risk behavior of the investors only under all conditions. Whether the formula in Equation 25 implies the risk behavior of investors may be open to further research.

Hence, it can be concluded that when the mathematical form of the utility curve of the payoff of the lottery, that is, the mathematical form of the utility curve of the uncertain value changes, the implications of $\rho = -\frac{u''(w)}{u'(w)}w$ may change as well.

If the investor has a fair gamble,

$$\rho = \frac{2\pi w_s}{\sigma_{\tilde{z}}^2} \tag{28}$$

holds true.

If the utility of the uncertain wealth value has a mathematical form such as in Figure 3,

$$\rho = \frac{2\pi w_s}{\beta \eta_t (E\tilde{z})^2} + \frac{2w_s}{E\tilde{z}} + \frac{(\beta \eta_t - 1)2u(w_s)w_s}{\beta \eta_t u'(w_s)(E\tilde{z})^2} \tag{29}$$

holds true.

When the utility of the uncertain wealth value is in of the form $\mathrm{E}u\ (w_{ns})$,

$$\rho = \frac{2\pi w_s}{\sigma_{\tilde{z}}^2} \tag{30}$$

holds true.

When the utility of the uncertain wealth value is of the form $\beta \eta_t u(w_s + E\tilde{z})$,

$$\rho = \frac{2\pi w_s}{\beta \eta_t (E\tilde{z})^2} + \frac{2w_s}{E\tilde{z}} + \frac{(\beta \eta_t - 1)2u(w_s)w_s}{\beta \eta_t u'(w_s)(E\tilde{z})^2} \tag{31}$$

holds true.

As a result, it can be concluded that $\rho = -\frac{u''(w)}{u'(w)}w$ in Equation 25 includes more than one component and does not only imply the approximate risk premium per variance of the gamble. Hence, $\rho = -\frac{u''(w)}{u'(w)}w$ may not be an appropriate coefficient in the study of risk behavior of investors under all conditions. Therefore, a new tool, in the form of the sufficiency factor of the model, must be developed to classify investors as risk-averse, risk-loving, and risk-neutral.

### 3.3 Solution to the Equity Premium Puzzle with the Sufficiency Factor of the Model

In this section, a new model has been developed to solve the equity premium puzzle after modifying the following equations from prior studies:

$$max_{\{\theta\}} u(c_t) + E_t[\beta u(c_{t+1})]$$

$$\text{s.t.} \tag{32}$$

$$c_t = e_t - p_t \theta$$

$$c_{t+1} = e_{t+1} + d\theta$$

(Cochrane, 2009, p. 5),

$$p_t u'(c_t) = \beta E_t[(p_{t+1} + y_{t+1})u'(c_{t+1})] \tag{33}$$

(fundamental pricing relationship) (Cochrane, 2009, p. 5),

$$ln\, E_t\,(R_{e,t+1}) = -ln\, \beta + \rho \mu_x - \frac{1}{2} \rho^2 \sigma_x{}^2 + \rho \sigma_{x,z} \tag{34}$$

(Mehra, 2008, p. 19),

$$ln\, R_f = -ln\,\beta + \rho \mu_x - \frac{1}{2} \rho^2 \sigma_x{}^2 \tag{35}$$

(Mehra, 2008, p. 19),

$$ln\, E(R_e) - ln\, R_f = \rho \sigma_x{}^2 \tag{36}$$

(Mehra, 2008, p. 19),

$$E\,(R_e) = \frac{E(x_{t+1})}{\beta E(x_{t+1}{}^{1-\rho})} \tag{37}$$

(Danthine & Donaldson, 2014, p. 291).

### 3.3.1 New Model

The risk behaviors of investors in the financial markets will be included in prior studies through the sufficiency factor of the model.

The representative agent maximizes his or her expected utility by

$$u(c_0) + \eta_t E_0 [\sum_{t=0}^{\infty} \beta^{t+1} u(c_{t+1})] \tag{38}$$

Here, $\eta_t$ is the sufficiency factor of the model that is determined at time t and it is in the form of coefficient. $E_0$ is the expectation operator conditional on information available at the present time 0. β denotes the subjective time discount factor that is equal to 0.99 (Danthine & Donaldson, 2014), $u$ is an increasing, continuously differentiable concave utility function, and $c$ shows the per capita consumption.

Hence, the problem faced by a typical investor may be explained with the inclusion of the sufficiency factor of the model in Equation 32 as follows:

$$max_{\{\theta\}} u(c_t) + \beta \eta_t E_t [u(c_{t+1})]$$

$$\text{s.t.} \tag{39}$$

$$c_t = e_t - p_t \theta$$

$$c_{t+1} = e_{t+1} + d\theta$$

Here,

$c_t$ and $c_{t+1}$      = per capita consumptions at time t and t+1, respectively,

$u$      = an increasing, continuously differentiable concave utility function,

$\beta$      = subjective time discount factor,

$\eta_t$      = sufficiency factor of the model that is determined at time t,

$e_t$ and $e_{t+1}$      = original consumption levels at time t and t+1, respectively if the investor

did not buy any asset,

$p_t$                             = price of the asset at time t,

$d$                              = payoff that is equal to $[p_{t+1} + y_{t+1}]$ in equity

and 1 under risk-free asset.

$\theta$                             = amount of asset

Alternative solution for the above problem of the typical investor by the dynamic optimization is also provided in the alternative derivation of Equation 40 in Appendix.

When a decision is made for financial assets, the utility curve of the uncertain financial assets may shift upward or downward automatically, according to the risk preferences of the investor. Hence, the sufficiency factor of the model is included in the problem of a typical investor because per capita consumptions include equity and risk-free assets.

The sufficiency factor of the model exists for the investor investing in risk-free asset because it is possible for the investor to sell the risk-free asset before maturity date (i.e., the investor may sell this asset to the Fed in the open market). As it will be shown in subsequent pages, there cannot exist no-trade equilibrium for the risk-free asset. This leads a possibility for the investor that the utility gained from 1 at maturity date may be different from the utility gained from the uncertain payoff of the risk-free asset that is decided to be sold before maturity date. For example, suppose that risk-free investor bought the risk-free asset at $95. The price of the asset increased to $98 before its maturity date and the investor decided to sell the asset at this price because of the macroeconomic conditions, the investor risk tolerance or the monetary policy of the Fed. The sufficiency factor of the model exists for the asset because $E_t(u(\$100))$ (i.e., predicted utility gained from certain face value at maturity) will be different from $E_t(u(\$98))$ (i.e., predicted utility gained from the uncertain $98). A future uncertainty for the investor will emerge at the beginning of the period because of his/her probable decision that will take place before the maturity date.

The alternative solution for the problem of the typical investor by the dynamic optimization has a binding constraint as $c_{t+1} = \theta_{t+1}y_{t+1} + \theta_{t+1}p_{t+1} - \theta_{t+2}p_{t+1}$ (see alternative derivation of Equation 40 in Appendix). Hence per capita consumption at time t+1, that is, $(c_{t+1})$ is equal to $\theta_{t+1}y_{t+1} + \theta_{t+1}p_{t+1} - \theta_{t+2}p_{t+1}$ that is uncertain from the view of point of investor who compares the utilities of certain and uncertain financial assets at time t . Hence the sufficiency factor of the model will exist for $E_t[u(c_{t+1})]$.

The economy is assumed to be frictionless.

Growth in per capita consumption is assumed to follow the Markov process. Moreover, one productive unit is assumed to produce output $y_t$ in period t, which is the period dividend. There exists one equity share that is competitively traded. Hence, the following formulas hold true for the following assumptions and definitions of the new model:

1. $u(c, \rho) = \frac{c^{1-\rho}}{1-\rho}$ ,

2. $c_t$ denotes per capita consumption,

3. $R_{e,t+1} = \frac{p_{t+1} + y_{t+1}}{p_t}$, where $p_{t+1}$ and $y_{t+1}$ are prices of the bond and dividends paid at time t+1, respectively,

4. $R_{f,t+1} = \frac{1}{q_t}$, where $q_t$ is the price of the bond,

5. the growth rate of consumption, $x_{t+1} = \frac{c_{t+1}}{c_t}$, is identically and independently distributed and log-normal,

6. the growth rate of dividends, $z_{t+1} = \frac{y_{t+1}}{y_t}$, is identically and independently distributed and log-normal,

7. return on equity is perfectly correlated with the growth rate of consumption (model equilibrium condition that $x = z$),

8. $(x_t, z_t)$ are jointly lognormally distributed,

9. gross return on equity $(R_{e,t})$ is identically and independently distributed $((R_{e,t})$ is identically and independently distributed because of the consequence of the above assumptions),

10. $(x_t, R_{e,t})$ are jointly lognormally distributed (because of the consequence of the above assumptions).

$$p_t u'(c_t) = \beta \eta_t E_t[(p_{t+1} + y_{t+1})u'(c_{t+1})] \quad \text{(fundamental pricing relationship)}, \tag{40}$$

See Derivation of Equation 40 in Appendix for proof.

$$ln\, E_t\,(R_{e,t+1}) = -ln\,\beta - ln\,\zeta_t + \rho\mu_x - \frac{1}{2}\rho^2\sigma_x{}^2 + \rho\sigma_{x,z}\,, \tag{41}$$

See Derivation of Equation 41 to 43 in Appendix for proof.

$$ln\, R_f = -ln\,\beta - ln\,\xi_t + \rho\mu_x - \frac{1}{2}\rho^2\sigma_x{}^2, \tag{42}$$

See Derivation of Equation 41 to 43 in Appendix for proof.

$$ln\,E(R_e)\, - ln\, R_f = ln\,\xi_t - ln\,\zeta_t + \rho\sigma_x{}^2, \tag{43}$$

See Derivation of Equation 41 to 43 in Appendix for proof.

$$E\,(R_e) = \frac{E(x_{t+1})}{\beta\zeta_t E(x_{t+1}{}^{1-\rho})}, \tag{44}$$

See Derivation of Equation 44 and Equation 45 in Appendix for proof.

$$ln\, E(R_e) = ln\, E(x_{t+1}) - ln\,\beta - ln\zeta_t - (1-\rho)\mu_x - \frac{1}{2}(1-\rho)^2\sigma_x{}^2 \tag{45}$$

See Derivation of Equation 44 and Equation 45 in Appendix for proof.

Here,

$\mu_x = E\,(ln\,x)$

$\sigma_x{}^2 = var\,(ln\,x)$

$\sigma_{x,z} = cov\,(ln\,x,\,ln\,z)$

$E(R_e) = $ mean equity for the period

$R_f = $ mean risk-free rate for the period

$ln\,E(R_e) = $ ln of mean equity

$ln\, R_f = $ ln of mean risk-free rate

$ln\,x = $ continuously compounded growth rate of consumption

$ln\ z$ = continuously compounded growth rate of dividend

$\beta$ = subjective time discount factor that is taken as 0.99

$\eta_t$ = sufficieny factor of the model determined at time t

$\zeta_t$ = sufficiency factor of the model determined at time t for the investors investing in equity

$\xi_t$ = sufficiency factor of the model determined at time t for the investors investing in risk-

free asset.

For the whole economy to be in equilibrium, the following must hold true:

1-$\theta_t = \theta_{t+1}$ =…= 1 exists for the equity investor. This means that the representative agent possesses the entire equity share;

2- the possession of the entire equity share entitles the representative agent to all the economy's dividend, that is, $c_t = y_t$ (Danthine &Donaldson, 2014, p.277);

3-No-trade equilibrium does not exist for the risk-free asset because Fed is able to sell or buy the risk-free asset in the open market because of its monetary policy aims from the representative agent.

Non-existence of the no-trade equilibrium for the risk-free asset can also be observed mathematically. Since $u(c_t) = \frac{c_t^{1-\rho}}{1-\rho}$ is selected for the utility curve of the investors, $\lim_{c_t \to 0} u'(c_t) = \infty$ holds true. This selection ensures that it is never optimal for the investor to choose a zero consumption level (Danthine &Donaldson, 2014, p.276).

When the fundamental equations of CCAPM are derived, the constraint may alternatively be selected as $\theta_{t+1}p_t + c_t \leq \theta_t d_t + \theta_t p_t$. Since the maximization of the objective function implies that the budget constraint will be binding, the constraint changes to $\theta_{t+1}p_t + c_t = \theta_t d_t + \theta_t p_t$.

If no-trade equilibrium for the risk-free asset is assumed, $\theta_t = \theta_{t+1}$ =…= 1 holds true in the equilibrium. This implies that the equilibrium value of consumption is equal to zero from the budget constraint of risk-free asset that is $\theta_{t+1}p_t + c_t = \theta_t p_t$. This value is obviously never optimal for the investor. Hence, $\theta_t = \theta_{t+1} = \ldots = 1$ does not hold true for the risk-free investor in the equilibrium.

4-with the inclusion of sufficiency factor of the model of the investor, the equilibrium price must satisfy Equation 40.

Equations 40 to 43 have been developed from the Equations 33 to 36, respectively by including the sufficiency factor of the model in Equations 33 to 36. Equation 45 is developed from Equation 37 by including the sufficiency factor of the model of the investors investing in equity.

The values of the sufficiency factor of the model for the investors investing in equity, the sufficiency factor of the model for the investors investing in risk-free asset, and the coefficient of relative risk aversion for the US economy for the period of 1889-1978 have been calculated from Equations 42, Equation 43, and Equation 45.

When $\mu_x$ and $\sigma_x^2$ that are calculated according to Table 1 are substituted in Equation 42, Equation 43, and Equation 45, the following system of equations will be generated:

$$-\ln(\zeta_t) - 0.017215(1-\rho) - \frac{1}{2}(1-\rho)^2 0.001250 = 0.039582, \qquad (46)$$

$$-\ln(\xi_t) + 0.017215\rho - 0.000625\rho^2 = \text{-}0.002082, \qquad (47)$$

$$\ln(\xi_t) - \ln(\zeta_t) + 0.001250\rho = 0.059504. \qquad (48)$$

[Insert table 1 here] The solution of the above system of equations with the sum of square errors (SSE) being $2.8\text{x}10^{-33}$ is:

$$\zeta_t \cong 0.961745$$

$$\xi_t \cong 1.019392$$

$$\rho \cong 1.033526.$$

Here, t denotes year 1977. The solution of the above system of equations was found with the NLSOLVE spreadsheet solver function of ExcelWorks LLC.

Subjective time discount factor is assigned a value as 0.99 for all calculations for the period. Input values of the equations 46 to 48 are taken with six decimal place accuracy because SSE of the solution of the above system of equations is the smallest at this accuracy.

## 4. Results and discussion

The newly derived model takes the subjective time discount factor as 0.99 and assigns the coefficient of relative risk aversion a value as 1.033526. Since these values are compatible with the empirical studies, they confirm that the model in this study provides a solution to the equity premium puzzle.

Since year 1978 was selected as the year of study for the determination of the risk behavior of investors, the solution of the above system of equations implies that as investors investing in risk-free asset increase their negative utility on the uncertain wealth value, while those investing in equity decrease their negative utility on the uncertain wealth value in year 1977. In other words, as investors investing in risk-free asset allocate negative utility for the uncertain wealth value, those investing in equity allocate utility for the uncertain wealth value in year 1977, with the value of the coefficient of relative risk aversion being 1.033526.

To detect the types of investors who invested in 1978, Equation 1, Equation 5, Equation 6, Equation 7, Equation 8, and Equation 9 should be referred to.

It has also been demonstrated with Equation 25 that the coefficient of relative risk aversion may not only imply the risk behavior of investors under all conditions. Since this coefficient may have three components, it can be concluded that the magnitude of the coefficient of relative risk aversion may not classify investors as risk-averse, risk-loving, or risk-neutral under all conditions.

The definitions for risk-averse, risk-loving, and risk-neutral investors have been reformulated and some new definitions for risk behavior of investors have been given.

This study certainly makes a significant contribution to the literature because the puzzle remains unsolved until date. Further, a new tool, that is, the sufficiency factor of the model has been used to classify investors as risk-averse, risk-loving, and risk-neutral in this study.

## 6. Conclusions

This study provides a solution to equity premium puzzle. The calculations of this new model show that the value of the coefficient of relative risk aversion is 1.033526 by assuming the value of the subjective time discount factor as 0.99. These values are found to be compatible with the empirical studies, confirming the validity of the derived model.

It is also shown with Equation 25 that the coefficient of relative risk aversion may not imply risk behavior of individuals under all conditions.

**Table 1**

*Statistics for the U.S. Economy for the period of 1889-1978*

| Statistics | Value |
| --- | --- |
| Mean return on equity $E(R_e)$ | 1.0698 |
| Mean risk-free rate, $R_f$ | 1.008 |
| Mean growth rate of consumption, $E(x)$ | 1.018 |
| Standard deviation of growth rate of consumption, $\sigma_x$ | 0.036 |
| Autocorrelation of consumption | -0.14 |
| Mean equity premium, $E(R_e) - R_f$ | 0.0618 |

Source: Adapted from *Handbook of the equity risk premium* (pp.19-20), by R. Mehra (2008).

Copyright 2008 by Elsevier.

**Figure 1**

*Fair gamble*

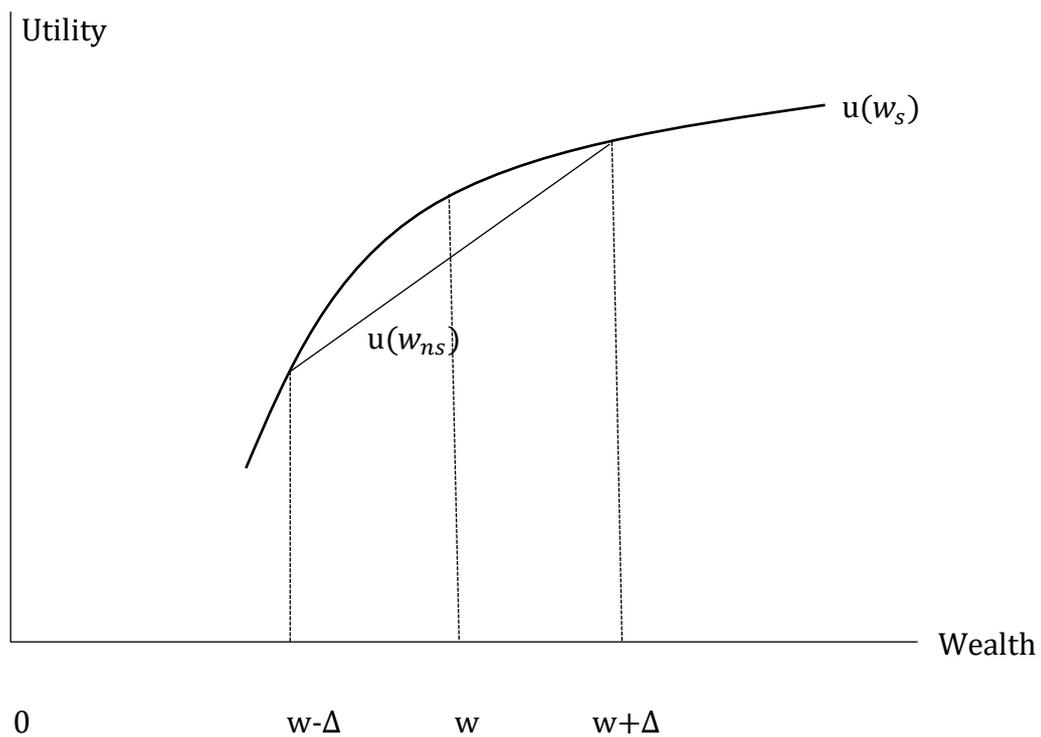

*Note.* The figure demonstrates the utility curves of the certain and uncertain wealth values. The utility curve of the certain wealth value $u(w_s)$ and the utility curve of the uncertain wealth value $u(w_{ns})$ intersect at the uncertain wealth values $w$-$\Delta$ and $w$+$\Delta$ that are at an equal distance from the certain wealth value ($w$).

**Figure 2**

*Utility curves of the certain and uncertain wealth values of the investor*

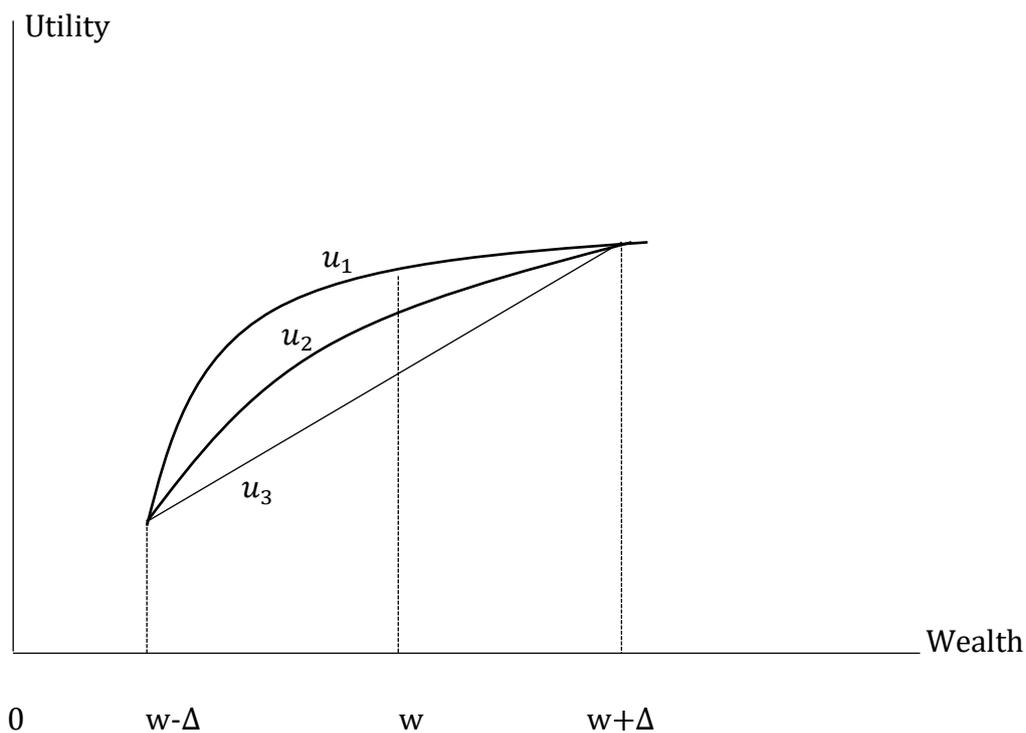

*Note.* The utility curve of the certain wealth value and the utility curve of the uncertain wealth value intersect at the uncertain wealth values *w-Δ* and w+Δ that are at equal distance from the certain wealth value *w*. As $u_1$ denotes the utility curve of the certain wealth value, $u_2$ and $u_3$ show the utility curves of the uncertain wealth values. As $u_3$ is the utility curve of the uncertain wealth value that is used in deriving $\rho = \frac{2\pi w_s}{\sigma_{\hat{z}}^2}$, $u_2$ may be an alternative utility curve of the uncertain wealth value.

**Figure 3**

*Risk premium graph for the investor allocating negative utility for the uncertain wealth value*

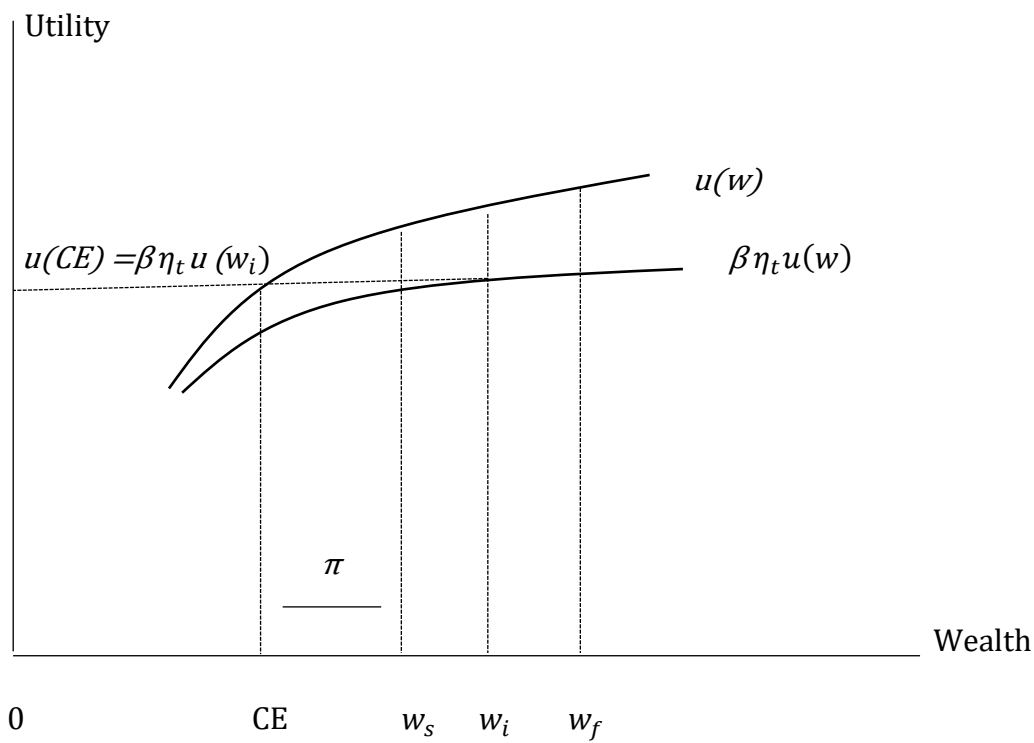

*Note.* The figure demonstrates that the investor expects $w_s$ to increase to $w_i$. The utility curve of the certain wealth value and the utility curve of the uncertain wealth value do not intersect at the uncertain wealth values that are at equal distance from the certain wealth value $w_s$. As CE is the certainty equivalent value, $\pi$ is the risk premium value. $u(CE) = \beta\eta_t u(w_i)$ holds true according to the figure.

**Figure 4**

*Graph of risk premium for the investor allocating the constant negative utility* $[-\delta_t]$ *for the uncertain wealth value*

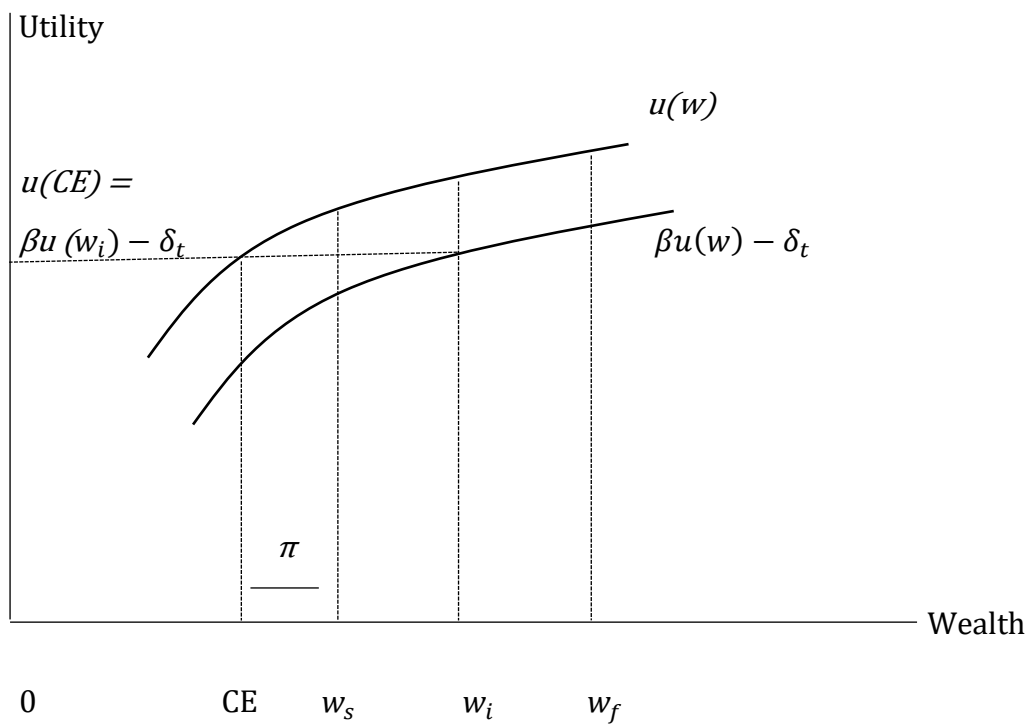

*Note.* The figure demonstrates that the investor allocates the constant negative utility for all wealth values for the uncertain wealth value. As CE is the certainty equivalent value, $\pi$ is the risk premium value. $u(CE) = \beta u\,(w_i) - \delta_t$ holds true according to the figure.

**Figure 5**

*Utility curves of the investor with ρ and η  values*

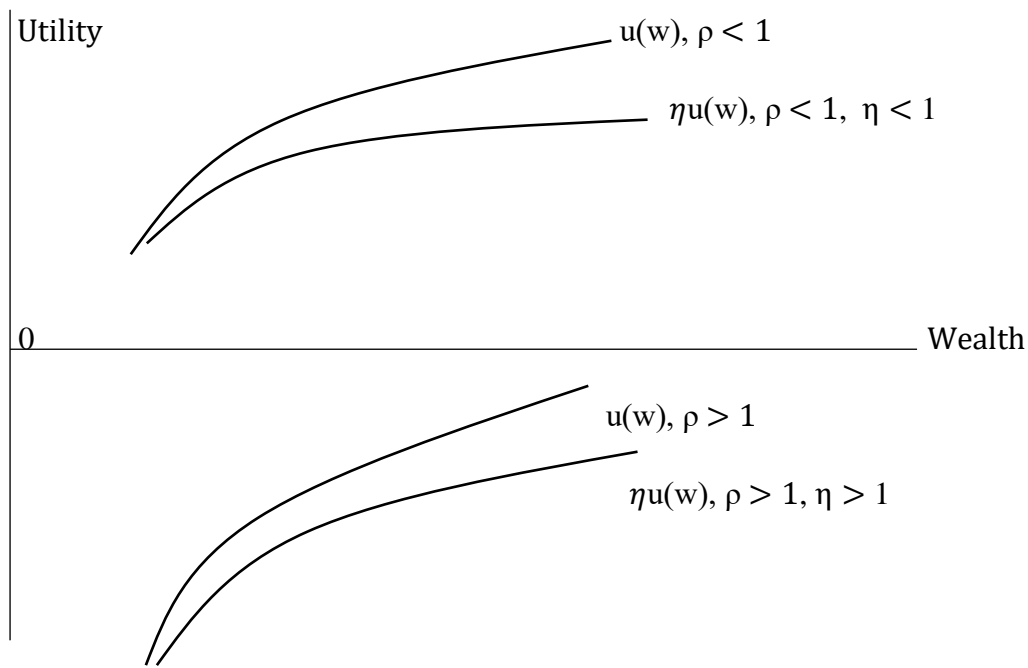

*Note.* The figure demonstrates the utility curves of an investor who allocates negative utility for the uncertain wealth value with $\rho$ and $\eta$ values where $\eta$ is the sufficiency factor of the model and $\rho$ is the coefficient of relative risk aversion.

**Figure 6**

*Utility curves of the investor with ρ and η  values*

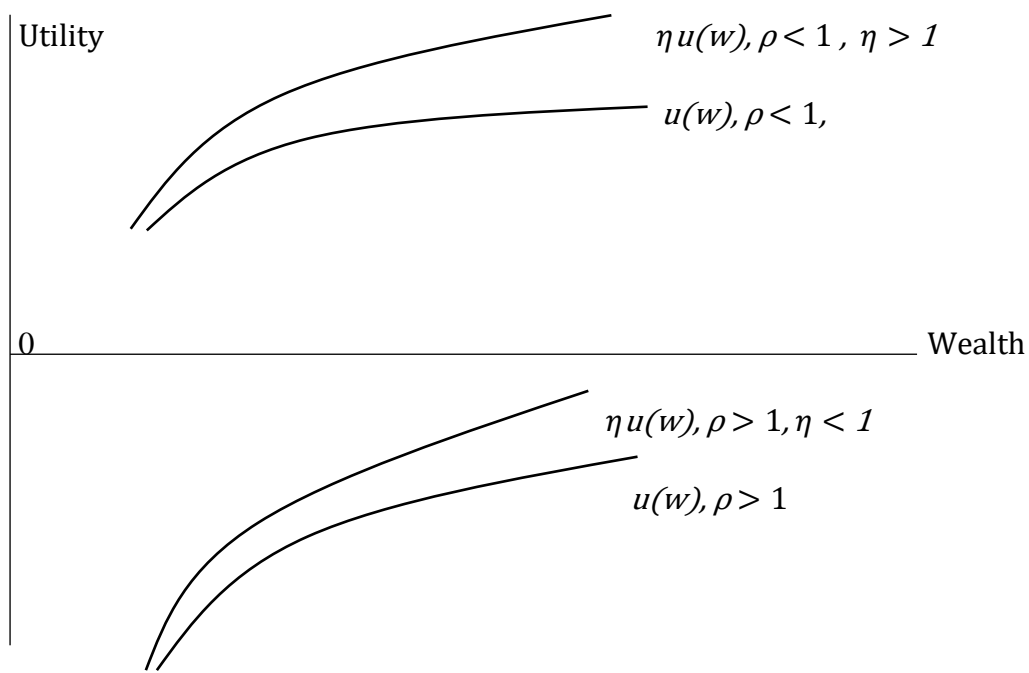

*Note.* The figure shows the utility curves of an investor who allocates utility for the uncertain wealth value with $\rho$ and $\eta$ values where $\eta$ is the sufficiency factor of the model and $\rho$ is the coefficient of relative risk aversion.

**Figure 7**

*Graph of risk premium for the investor who allocates negative utility for the uncertain wealth value*

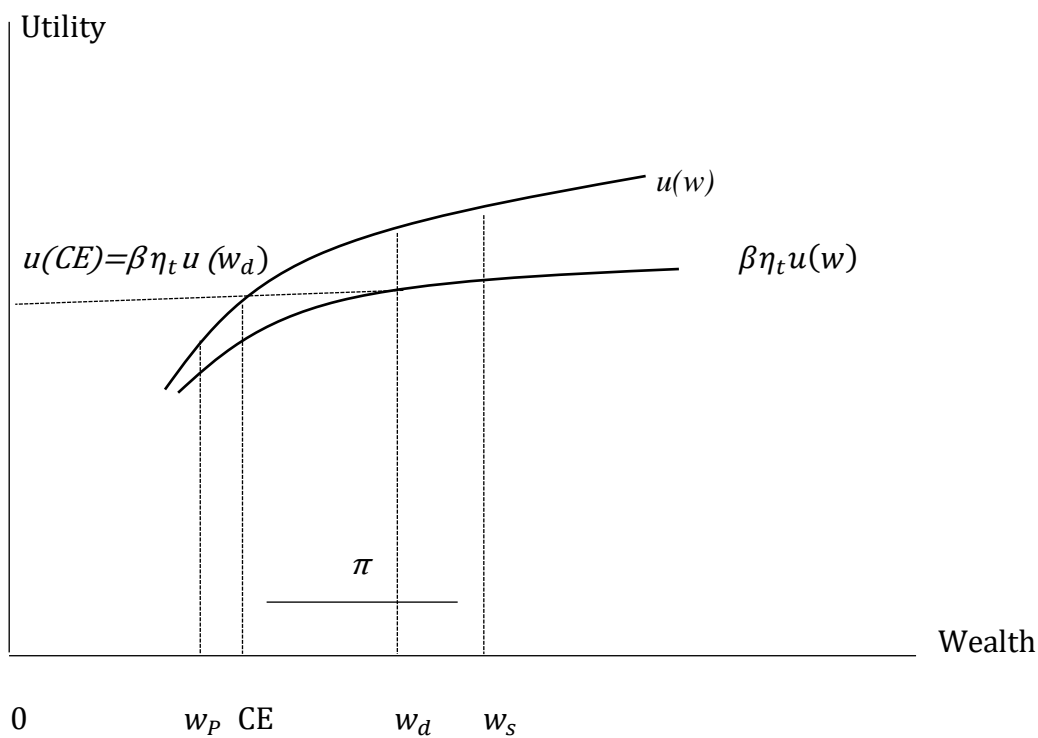

*Note.* The figure shows that investor expects $w_s$ to decrease to $w_d$. The utility curve of the certain wealth value and the utility curve of the uncertain wealth value do not intersect at the uncertain wealth values that are at equal distance from the sure wealth value $w_s$. As CE is the certainty equivalent value, $\pi$ is the risk premium value. $u(CE) = \beta \eta_t u(w_d)$ holds true according to the figure.

**Appendix**

**Derivation of Equation 27**

If the left-hand side of Equation 18 is expanded around $w_s$ and is made equal to

$$\beta \eta_t u(w_{ns}),$$ (A.1)

we have

$$u(w_s) - \pi u'(w_s) + O(\pi^2) = \beta \eta_t u(w_{ns}).$$ (A.2)

Algebraically,

$$\pi \cong -\frac{\beta \eta_t u(w_{ns}) - u(w_s)}{u'(w_s)}.$$ (A.3)

Now, if both sides of Equation A.3 are multiplied by $[-u''(w_s)]$, after algebraic operations we have

$$\pi \cong \frac{\rho[\eta_t \beta u(w_{ns}) - u(w_s)]}{u''(w_s)}.$$ (A.4)

**Derivation of Equation 40** (Cochrane, 2009, p. 5)

Now, we place the constraints in the objective function and set the derivative with respect to $\theta$ equal to zero in Equation 39 with $d = p_{t+1} + y_{t+1}$, we have Equation 40.

**Alternative Derivation of Equation 40 by Dynamic Optimization**

The problem of typical investor will be

$$max_{\{\theta_{t+1}, c_t\}} \{u(c_0) + \zeta_t E_0[\sum_{t=0}^{\infty} \beta^{t+1} u(c_{t+1})]\}$$

s.t. (A.5)

$$\theta_{t+1} p_t + c_t \leq \theta_t y_t + \theta_t p_t$$

where 0 and $\zeta_t$ denote present time and the sufficiency factor of the model of the equity investors, respectively.

We will use Bellman's optimality principle to solve the problem. The Bellman's optimality principle provides us to write the problem in A.5 as the following two period one:

$$V_t(\theta_t) = max_{\{\theta_{t+1}, \ c_t\}} \{u(\ c_t) + \zeta_t \beta E_t[\ V_{t+1}(\theta_{t+1})]\} \tag{A.6}$$

$$\text{s.t.}$$

$$\theta_{t+1} = \frac{(y_t + p_t)\theta_t - c_t}{p_t} \ . \tag{A.7}$$

We will write A.6 and A.7 in Bellman form to derive the first order conditions by using Lagrangian as follows:

$$V_t(\theta_t) = max_{\{c_t, \ \theta_{t+1}\}} [u(\ c_t) + \zeta_t \beta E_t(\ V_{t+1}(\theta_{t+1}))] + \lambda_t[\theta_{t+1}p_t \ + \ c_t - \theta_t y_t - \theta_t p_t \ ] \ . \tag{A.8}$$

Differentiate with respect to $c_t$ and $\theta_{t+1}$ in A.9 and A.10, respectively get the first order conditions:

$$u'(\ c_t) + \lambda_t = 0 \tag{A.9}$$

$$\zeta_t \beta E_t \ [\ V'_{t+1} \ (\theta_{t+1})] + \lambda_t p_t = 0 \ . \tag{A.10}$$

Hence, from A.9 and A.10

$$u'(\ c_t) = \zeta_t \beta E_t \ [\ V'_{t+1} \ (\theta_{t+1})] \frac{1}{p_t} \ . \tag{A.11}$$

To get the envelope condition, just take the derivative of A.8 with respect to $\theta_t$:

$$V'_t(\theta_t) = -\lambda_t \ (y_t + p_t). \tag{A.12}$$

Shift up one period, we get

$$V'_{t+1}(\theta_{t+1}) = -\lambda_{t+1} \ (y_{t+1} + p_{t+1}) \tag{A.13}$$

and

$$- \lambda_{t+1} = u'(\ c_{t+1}) \tag{A.14}$$

holds true according to A.9.

Substitute A.14 in A.13 to get

$$V'_{t+1}(\theta_{t+1}) = u'(c_{t+1})(y_{t+1} + p_{t+1}).$$ (A.15)

Substitute A.15 in A.11 to get

$$p_t u'(c_t) = \zeta_t \beta E_t[u'(c_{t+1})(p_{t+1} + y_{t+1})].$$ (A.16)

**Derivation of Equation 41 to 43** (Mehra, 2008, pp. 17-19)

Since the equity price is homogeneous of degree 1 in y, the equity price is in the form of

$$p_t = vy_t,$$ (A.17)

where v is a constant coefficient.

By the fundamental pricing relationship, we have

$$vy_t = \beta \zeta_t E_t[(vy_{t+1} + y_{t+1}) \frac{u'(c_{t+1})}{u'(c_t)}].$$ (A.18)

Hence,

$$V = \frac{\beta \zeta_t E_t(z_{t+1} x_{t+1}^{-\rho})}{1 - \beta \zeta_t E_t(z_{t+1} x_{t+1}^{-\rho})}.$$ (A.19)

By definition,

$$R_{e,t+1} = \frac{p_{t+1} + y_{t+1}}{p_t}$$ (A.20)

or,

$$R_{e,t+1} = (\frac{v+1}{v})(\frac{y_{t+1}}{y_t}) = (\frac{v+1}{v}) z_{t+1}.$$ (A.21)

Taking conditional expectation on both sides of Equation A.21 results in

$$E_t(R_{e,t+1}) = (\frac{v+1}{v}) E_t(z_{t+1}).$$ (A.22)

Substituting

$$\left(\frac{v+1}{v}\right) = \frac{1}{\beta\,\zeta_t E_t(z_{t+1} x_{t+1}^{-\rho})} \tag{A.23}$$

in $E_t(R_{e,t+1})$ results in

$$\frac{E_t(z_{t+1})}{\beta\,\zeta_t E_t(z_{t+1} x_{t+1}^{-\rho})} \ . \tag{A.24}$$

Similarly,

$$R_{f,t+1} = \frac{1}{\beta\xi_t\,E_t(x_{t+1}^{-\rho})} \ . \tag{A.25}$$

By using the lognormal properties, we have

$$E_t(R_{e,t+1}) = \frac{exp(\mu_z + \frac{1}{2}\sigma_z{}^2)}{\beta\zeta_t exp\,[\mu_z - \rho\mu_x + \frac{1}{2}(\sigma_z{}^2 + \rho^2\sigma_x{}^2 - 2\,\rho\sigma_{x,z})]} \tag{A.26}$$

and

$$R_f = \frac{1}{\beta\xi_t\,exp\,(-\rho\mu_x + \frac{1}{2}\rho^2\sigma_x{}^2)} \ . \tag{A.27}$$

If we take ln on both sides, we get

$$ln\,E_t\,(R_{e,t+1}) = -ln\,\beta - ln\,\zeta_t + \rho\mu_x - \frac{1}{2}\rho^2\sigma_x{}^2 + \rho\sigma_{x,z} \tag{A.28}$$

and

$$ln\,R_f = -ln\,\beta - ln\,\xi_t + \rho\mu_x - \frac{1}{2}\rho^2\sigma_x{}^2. \tag{A.29}$$

Since $R_{e,t}$ is identically and independently distributed,

$$ln\,E_t\,(R_{e,t+1}) = ln\,E\,(R_e) = -ln\,\beta - ln\,\zeta_t + \rho\mu_x - \frac{1}{2}\rho^2\sigma_x{}^2 + \rho\sigma_{x,z} \ . \tag{A.30}$$

Subtracting $ln\,R_f$ from $ln\,E\,(R_e)$ results in

$$ln\,E\,(R_e) - ln\,R_f = ln\,\xi_t - ln\,\zeta_t + \rho\sigma_{x,z} \ . \tag{A.31}$$

From Equation A.21 we also get

$$ln\, E\,(R_e) - ln\, R_f = ln\, \xi_t - ln\, \zeta_t + \ \rho \sigma_{x,R_e} \,, \tag{A.32}$$

where

$$\sigma_{x,R_e} = cov\,(ln\, x,\, ln\, R_e). \tag{A.33}$$

Since the equilibrium condition sets x = z, we have

$$ln\, E\,(R_e) - ln\, R_f = ln\, \xi_t - ln\, \zeta_t + \ \rho \sigma_x{}^2. \tag{A.34}$$

**Derivation of Equation 44 and Equation 45** (Danthine & Donaldson, 2014, pp. 289-291)

The equity price is of the form

$$p_t = v y_t \,, \tag{A.35}$$

where v is a constant coefficient.

From Equation 40 we have

$$v y_t = \beta \ \zeta_t E_t \big[ (v y_{t+1} + y_{t+1}) \tfrac{u'(c_{t+1})}{u'(c_t)} \big]. \tag{A.36}$$

Since $x_{t+1}$ is identically and independently distributed,

$$v y_t = \beta \zeta_t E \big[ (v + 1) \tfrac{y_{t+1}}{y_t} x_{t+1}{}^{-\rho} \big]. \tag{A.37}$$

The market clearing condition requires that

$$\tfrac{y_{t+1}}{y_t} = x_{t+1} \,. \tag{A.38}$$

Therefore,

$$v = \beta \zeta_t E \big[ (v + 1) \, x_{t+1}{}^{1-\rho} \big] \tag{A.39}$$

or,

$$v = \tfrac{\beta \zeta_t E( x_{t+1}{}^{1-\rho})}{1 - \beta \zeta_t E( x_{t+1}{}^{1-\rho})} \,. \tag{A.40}$$

Hence, v is indeed a constant.

Since

$$R_{e,t+1} = \frac{p_{t+1} + y_{t+1}}{p_t} = \frac{v+1}{v} \frac{y_{t+1}}{y_t} = \frac{v+1}{v} x_{t+1} \ , \tag{A.41}$$

the following holds true if expectations are taken on both sides of Equation A.41:

$$E_t(R_{e,t+1}) = E(R_e) = \frac{v+1}{v} E(x_{t+1}) = \frac{E(x_{t+1})}{\beta \zeta_t E(x_{t+1}^{1-\rho})} \tag{A.42}$$

or,

$$E_t(R_{e,t+1}) = E(R_e) = \frac{E(x_{t+1})}{\beta \zeta_t exp\left[(1-\rho)\,\mu_x + \frac{1}{2}(1-\rho)^2 \sigma_x^2)\right]} \ . \tag{A.43}$$

Taking ln on both sides results in

$$ln\,E(R_e) = ln\,E(x_{t+1}) - ln\ \beta\ - ln\,\zeta_t - (1\text{-}\rho)\,\mu_x - \frac{1}{2}(1-\rho)^2 \sigma_x^2. \tag{A.44}$$